\documentclass[letter,twocolumn]{jpsj3}
\usepackage{amsthm}
\usepackage{txfonts}
\usepackage{graphicx}
\usepackage{bm}
\usepackage{physics}
\usepackage{xcolor}
\providecommand{\RevisionMode}{0}
\newif\ifshowchanges
\ifnum\RevisionMode=1\relax\showchangestrue\else\showchangesfalse\fi
\DeclareRobustCommand{\rev}[1]{{\ifshowchanges\color{red}\else\color{black}\fi#1}}
\newenvironment{revised}{\begingroup\ifshowchanges\color{red}\else\color{black}\fi}{\endgroup}
\renewcommand{\footnoterule}{\noindent\vskip1\baselineskip\kern-3pt
  \hbox{\color{black}\vrule width\columnwidth height.4pt depth0pt}\kern2.6pt}

\newtheorem{proposition}{Proposition}

\title{Boundary-Cancelled Score-Hamiltonian Sampling}

\author{Masayuki Ohzeki$^{1,2,3,4}$\thanks{\color{black}Corresponding author. masayuki.ohzeki.a4@tohoku.ac.jp}}

\inst{
$^1$Graduate School of Information Sciences, Tohoku University, Sendai 980-8579, Japan\\
$^2$Department of Physics, Institute of Science Tokyo, Tokyo 152-8551, Japan\\
$^3$Research and Education Institute for Semiconductors and Informatics, Kumamoto University, Kumamoto 860-8555, Japan\\
$^4$Sigma-i Co., Ltd., Tokyo 108-0075, Japan
}

\abst{
\rev{Diffusion sampling is related to imaginary-time quantum annealing through a forward/backward Schr\"odinger equation pair and its one-sided counterdiabatic connection.
For a smooth, gapped Score-Hamiltonian path, the exact connection transports the target density at any evolution time.
We apply quantum-annealing performance analysis to residuals from approximating this connection with a trained score.
If the first $r$ terminal schedule derivatives vanish, the squared Hellinger error scales as $T^{-2r-2}$, with a prefactor determined by the terminal learning residual and inverse gaps.
Direct two-dimensional imaginary-time evolution with one fixed neural score reproduces these powers and prefactors.
Terminal schedule design thus suppresses the distributional effect of imperfect score learning without retraining the network.}
}

\begin{document}
\maketitle

{\it Introduction.---}
Diffusion models are usually introduced by a stochastic differential equation: a noising process sends a data distribution to a simple prior, and a reverse process generates data from that prior.\cite{Ho2020DDPM,Song2021ScoreSDE,Anderson1982}
To understand the extra force required for reversal, we describe the density path using two positive real amplitudes.
One obeys a diffusion-like, imaginary-time Schr\"odinger equation; the other obeys its formal adjoint, with the opposite sign of the time derivative.
Their product, rather than either amplitude alone, is the probability density, and their logarithmic gradients determine the forward and backward drifts.\cite{Nelson1966,Zambrini1987}
Here ``adjoint'' specifies a relation between evolution equations, not by itself a reversed stochastic process; the drift and time orientation must also be identified.
Schr\"odinger bridges use this two-amplitude structure to select a path law by minimizing relative entropy at fixed initial and final densities.\cite{Schrodinger1931,Leonard2014Survey,KabaShimizuOhzekiSughiyama2025}
Here the diffusion path is already given: its factorization exposes the two drifts without solving an endpoint-constrained entropy minimization.

\rev{Simulated annealing (SA) and quantum annealing (QA) provide computational routes from disordered states to meaningful order.\cite{KirkpatrickGelattVecchi1983,KadowakiNishimori1998}
Although commonly formulated as optimization methods, they extract information in the form of low-cost solutions by gradually reducing thermal or quantum fluctuations.
Diffusion models similarly generate structured data from noise, motivating a structural connection with annealing.
Here we develop this connection specifically to imaginary-time QA: the diffusion-like evolution $\partial_t\psi=-{\cal H}\psi$, also called Euclidean evolution, rather than the unitary real-time equation $i\partial_t\psi={\cal H}\psi$ (with $\hbar=1$).
Stella, Santoro, and Tosatti evaluated imaginary-time QA by directly integrating the Schr\"odinger equation and comparing residual-energy decay with real-time QA and classical Fokker--Planck annealing.\cite{StellaSantoroTosatti2005}
This connection motivates using annealing theory to uncover theoretical relations and identify ways to improve the fidelity of diffusion-based generation.}

The central observation of this Letter is that the score force needed for reverse diffusion has a counterdiabatic (CD) interpretation: it supplies the correction that transports the prescribed density path in the generation direction.
The square root of the density defines a ground state of a supersymmetric Score Hamiltonian; fixing the forward drift selects an adjoint amplitude whose motion yields the corresponding one-sided imaginary-time CD connection.
Recent work has already mapped score-based sampling to adiabatic transport in a Score Hamiltonian.\cite{HalmosHanin2026ScoreHamiltonian}
Counterdiabatic driving and geometric adiabatic transport are likewise well established.\cite{Kato1950,Berry2009,Jarzynski2013,SelsPolkovnikov2017}
\rev{We study the finite-time residual when a trained score supplies an imperfect CD connection along a prescribed Score-Hamiltonian path.
Terminal flattening suppresses this forcing through the Morita--Nishimori boundary-cancellation mechanism.\cite{Morita2007Faster,MoritaNishimori2008MathFoundation,VenutiLidar2018Boundary}
We derive the corresponding Hellinger asymptotics and test them by direct evolution with one frozen neural score.}

{\it Imaginary-time pair and drifts.---}
Let $\tilde\psi_t,\psi_t>0$ be auxiliary amplitudes for the same real, spatially self-adjoint operator
\begin{equation}
    {\cal H}_t=-D\nabla^2+V_t(x).
    \label{eq:euclidean-hamiltonian}
\end{equation}
\rev{Here $x\in\mathbb R^d$ is the spatial coordinate, $\nabla$ is its gradient, $D>0$ is the diffusion coefficient, and $V_t$ is a real potential.}
We take $\partial_t\psi_t=-{\cal H}_t\psi_t$ and its formal adjoint $\partial_t\tilde\psi_t={\cal H}_t\tilde\psi_t$.
The latter propagates terminal data toward earlier times; the sign change alone is not yet the drift rule for a reversed diffusion.
Neither amplitude is separately a normalized probability density.
Their product
\begin{equation}
    P_t(x)=\tilde\psi_t(x)\psi_t(x)
    \label{eq:product-density}
\end{equation}
is the time-dependent probability density after fixing its normalization.
The potential terms cancel on differentiating the product, giving
\begin{equation}
    \partial_tP_t
    =
    D\nabla\cdot\left(\tilde\psi_t\nabla\psi_t-\psi_t\nabla\tilde\psi_t\right).
    \label{eq:current-from-pair}
\end{equation}
With vanishing boundary flux, this equation conserves $\int P_t\,dx$.
It also identifies two stochastic descriptions of the same density path.
First, eliminating $\psi_t=P_t/\tilde\psi_t$ gives the forward Fokker--Planck equation
\begin{equation}
    \partial_tP_t=-\nabla\cdot(v_+P_t)+D\nabla^2P_t
    \label{eq:fp-forward}
\end{equation}
\rev{where $v_+=2D\nabla\log\tilde\psi_t$ is the forward drift.}
Second, eliminating $\tilde\psi_t=P_t/\psi_t$ rewrites the same equation as $\partial_tP_t=-\nabla\cdot(v_-P_t)-D\nabla^2P_t$, with the backward drift
\begin{equation}
    v_-=-2D\nabla\log\psi_t .
    \label{eq:fp-backward}
\end{equation}
\rev{In increasing reverse time $\tau=t_{\rm f}-t$, where $t_{\rm f}$ is the final noising time, the diffusion coefficient is again $D>0$ and the drift is $-v_-$.}
Thus these are forward and reversed processes, with the same densities visited in opposite order, not two independent density paths.
\rev{The density product alone does not determine the two amplitudes; choosing the forward drift fixes the factorization up to a time-dependent scalar.}
The difference of the two drift fields is fixed by the density:
\begin{equation}
    v_+-v_-=2D\nabla\log P_t .
    \label{eq:drift-difference}
\end{equation}
Consequently, the reversed drift is $-v_-=-v_++2D\nabla\log P_t$: the logarithmic force is the correction to simply reversing the forward drift.

{\it Supersymmetric Hamiltonian path.---}
Conversely, a desired density path can be turned into a ground-state path.
\rev{For a smooth normalized positive density path $\rho_\lambda$, with $0\leq\lambda\leq1$, set $\phi_\lambda=\sqrt{\rho_\lambda}$ and define ${\cal Q}_\lambda=\sqrt D(\nabla-\nabla\log\phi_\lambda)$.}
\rev{Adjoints and norms use the spatial $L^2$ inner product or its discrete counterpart.}
The supersymmetric Hamiltonian $\widehat H_\lambda={\cal Q}_\lambda^\dagger{\cal Q}_\lambda$ \rev{becomes}
\begin{equation}
    \widehat H_\lambda
    =
    -D\nabla^2
    +D\frac{\nabla^2\phi_\lambda}{\phi_\lambda}.
    \label{eq:susy-hamiltonian-phi}
\end{equation}
\rev{Then} $\widehat H_\lambda\phi_\lambda=0$.
\rev{Writing $S_\lambda=\nabla\log\rho_\lambda$ gives}
\begin{revised}
\begin{equation}
    \widehat H_\lambda=-D\nabla^2+D\left(\frac12\nabla\cdot S_\lambda+\frac14 |S_\lambda|^2\right).
    \label{eq:score-hamiltonian}
\end{equation}
\end{revised}
\begin{revised}
The potential in Eq.~(\ref{eq:score-hamiltonian}) also has a familiar machine-learning interpretation.
For a neural score $S_\lambda^\theta$ with parameters $\theta$, define the score-matching loss $\mathcal L_{\rm SM}(\theta)=(1/2)\mathbb E_{\rho_\lambda}|S_\lambda^\theta-S_\lambda|^2$, where $\mathbb E_{\rho_\lambda}$ denotes expectation under $\rho_\lambda$.
Integration by parts, with vanishing boundary term, gives\cite{Hyvarinen2005ScoreMatching}
\begin{equation}
    \mathcal L_{\rm SM}(\theta)
    =\frac{2}{D}\mathbb E_{\rho_\lambda}[V_\lambda^\theta]+c_\lambda^{\rm SM},
    \label{eq:score-matching-potential}
\end{equation}
where $V_\lambda^\theta=D[(1/2)\nabla\cdot S_\lambda^\theta+(1/4)|S_\lambda^\theta|^2]$ and $c_\lambda^{\rm SM}=(1/2)\mathbb E_{\rho_\lambda}|S_\lambda|^2$ is independent of $\theta$.
Thus $2V_\lambda^\theta/D$ is precisely the standard implicit score-matching integrand, $\nabla\cdot S_\lambda^\theta+(1/2)|S_\lambda^\theta|^2$.

Our purpose in introducing $\widehat H_\lambda$ is to express generation as transport of a ground state whose probability density is $\rho_\lambda$.
This formulation lets us ask how a finite evolution time and an imperfect transport force change the final generated distribution.
Imaginary-time annealing theory relates the surviving excited-state amplitudes to the derivatives of the schedule at the terminal time.
We use this relation to choose schedules whose terminal derivatives vanish, thereby suppressing the final distribution error.
The identity $\widehat H_\lambda\phi_\lambda=0$ fixes the instantaneous state, but transport also requires its velocity.
\end{revised}

{\it Forward and reverse roles.---}
The forward noising process is usually specified directly as a Fokker--Planck, or stochastic, dynamics.
Fix its drift $b_t$ by
\begin{equation}
    \partial_tP_t=-\nabla\cdot(b_tP_t)+D\nabla^2P_t .
    \label{eq:fp-with-b}
\end{equation}
Equation~(\ref{eq:fp-with-b}) is the chosen stochastic dynamics that moves $P_t$.
\rev{The instantaneous Hamiltonian reconstructed from $P_t$ by Eq.~(\ref{eq:susy-hamiltonian-phi}) records its ground state and spectral gaps; it is distinct from the Fokker--Planck generator.}
The imaginary-time amplitude dynamics of Eq.~(\ref{eq:euclidean-hamiltonian}) requires the additional factorization $P_t=\tilde\psi_t\psi_t$.
The time-reversed stochastic process is not obtained by only changing $b_t$ to $-b_t$.
When generation is parameterized in the increasing reverse time, the drift is
\begin{equation}
    b_{\rm gen}(x,t)=-b_t(x)+2D S_t(x)
    \label{eq:reverse-score-drift}
\end{equation}
where $S_t=\nabla\log P_t$.
This is the earlier relation $-v_-=-v_++2D\nabla\log P_t$ with $v_+=b_t$.
The additional force $2DS_t$ is the score correction: it is required to retrace the density path, rather than merely negate the noising force.
\rev{The same score enters the instantaneous Hamiltonian in Eq.~(\ref{eq:score-hamiltonian}).}
\rev{In the terminology of quantum annealing, the additional term that supplies the velocity of a moving instantaneous state is a CD correction.
To identify this correction for the score-driven reverse process, we now rewrite the density dynamics as an imaginary-time amplitude equation.}

For the gradient drifts considered here, choose $\tilde\psi_t$ so that $b_t=2D\nabla\log\tilde\psi_t$.
This fixes the forward noising gauge, or equivalently the adjoint amplitude, up to a time-dependent scalar.
The associated instantaneous supersymmetric operator is
\begin{equation}
    \widehat H_{\tilde\psi,t}
    =
    -D\nabla^2+D\frac{\nabla^2\tilde\psi_t}{\tilde\psi_t}.
    \label{eq:tilde-susy-hamiltonian}
\end{equation}
It satisfies $\widehat H_{\tilde\psi,t}\tilde\psi_t=0$ and records the zero mode fixed by $b_t$.
Substituting $P_t=\tilde\psi_t\psi_t$ and $b_t=2D\nabla\log\tilde\psi_t$ into Eq.~(\ref{eq:fp-with-b}), and then dividing by $\tilde\psi_t$, gives the equation for the complementary amplitude, with derivatives still taken in the original noising coordinate $t$ (in increasing reverse time, $\partial_\tau=-\partial_t$):
\begin{equation}
    \partial_t\psi_t
    =
    -\left(\widehat H_{\tilde\psi,t}+G^{\rm ECD}_t\right)\psi_t
    \label{eq:ecd-hamiltonian-psi}
\end{equation}
\rev{where $G^{\rm ECD}_t=\partial_t\log\tilde\psi_t$ is the connection of the adjoint amplitude, acting with the dual minus sign on $\psi_t$ in Eq.~(\ref{eq:ecd-hamiltonian-psi}); ECD denotes imaginary-time (Euclidean) CD.}
\rev{For square-integrable $\tilde\psi_t$, normalize the positive amplitude as $\ket{\tilde\phi_t}=\tilde\psi_t/\|\tilde\psi_t\|$ and set $\Pi_t=\ket{\tilde\phi_t}\bra{\tilde\phi_t}$.}
The action of $G^{\rm ECD}_t$ on $\ket{\tilde\phi_t}$ contains a component parallel to $\ket{\tilde\phi_t}$, which changes only the normalization or gauge of the imaginary-time amplitude.
The physical counterdiabatic part is the transverse component.
\rev{With $Q_t=1-\Pi_t$, and a dot denoting $\partial_t$, one obtains}
\begin{revised}
\begin{equation}
    Q_tG^{\rm ECD}_t\ket{\tilde\phi_t}
    = Q_t\ket{\dot{\tilde\phi}_t}.
    \label{eq:cd-transverse-action}
\end{equation}
\end{revised}
The projector $Q_t$ therefore extracts exactly the part that couples the instantaneous state to its orthogonal complement.
The Hermitian real-time counterdiabatic Hamiltonian associated with the same path is obtained by the off-diagonal lift
\begin{revised}
\begin{equation}
    H_{\rm CD}^{(\tilde\psi)}
    =i\left(Q_tG^{\rm ECD}_t\Pi_t-\Pi_tG^{\rm ECD}_tQ_t\right).
    \label{eq:cd-hermitian-lift}
\end{equation}
\end{revised}
This is the Kato--Berry expression $H_{\rm CD}^{(\tilde\psi)}=i[\dot\Pi_t,\Pi_t]$.
\begin{revised}
The connection can be read operationally: the vector drift $2DS_t$ transports the probability density in Eq.~(\ref{eq:reverse-score-drift}); rewriting $P_t=\tilde\psi_t\psi_t$ produces the multiplicative correction $G_t^{\rm ECD}$ in the equation for $\psi_t$.
Equation~(\ref{eq:cd-transverse-action}) identifies its transverse action with the velocity of the normalized adjoint state, and Eq.~(\ref{eq:cd-hermitian-lift}) gives the corresponding Hermitian CD operator.

The Score-Hamiltonian mapping and adiabatic sampling estimates are established,\cite{HalmosHanin2026ScoreHamiltonian} as are projector CD transport\cite{Kato1950,Berry2009} and terminal boundary cancellation.\cite{Morita2007Faster,MoritaNishimori2008MathFoundation,VenutiLidar2018Boundary}
Here we derive the one-sided amplitude connection from a fixed noising drift and its reverse score force, then apply terminal cancellation to the residual of learned CD transport along a prescribed ground-state path.
\end{revised}

\begin{figure*}[t]
    \centering
    \includegraphics[width=0.68\textwidth]{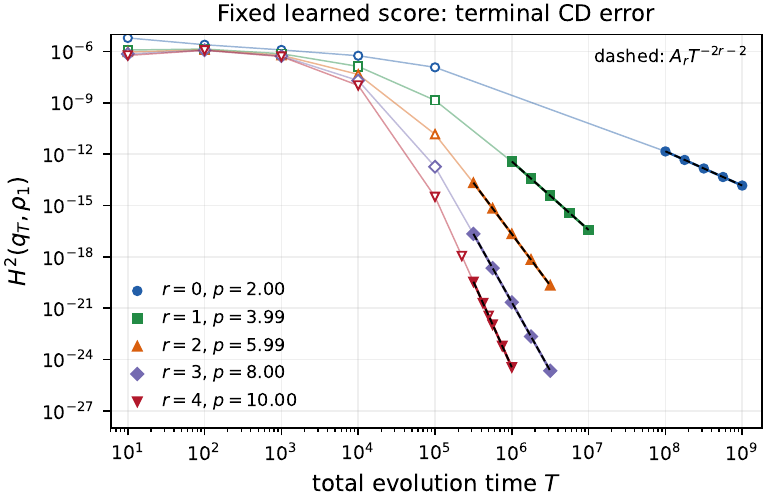}
    \caption{\rev{(Color online) Residual squared Hellinger error versus total evolution time $T$ for one fixed learned CD connection, with $r=0,\ldots,4$.
    Symbols are direct integrations of Eq.~(\ref{eq:learned-cd-evolution}) on a $49\times49$ spatial grid in $[-8,8]^2$; filled symbols are the five logarithmically spaced points used in each asymptotic fit.
    Colored segments fit only the prefactor at the predicted exponent $2r+2$; black dashed lines use the independent endpoint prefactor in Eq.~(\ref{eq:cd-endpoint-prefactor}).
    Legend values $p$ are free-exponent fits $H^2\propto T^{-p}$.
    Fit ranges are $[10^8,10^9]$, $[10^6,10^7]$, $[10^{5.5},10^{6.5}]$, $[10^{5.5},10^{6.5}]$, and $[10^{5.5},10^6]$ for increasing $r$.
    The noising rate is $\beta=8$, with $D=4$.
    Target mixture weights are $(.24,.31,.22,.23)$, means $(-3,-2),(2.7,-1.5),(-.9,2.5),(2.4,2.2)$, and component standard deviations along the two axes $(.46,.58),(.58,.46),(.70,.44),(.48,.64)$.
    A network with three hidden layers of 96 units and SiLU activation $z\mapsto z/(1+e^{-z})$ is trained for 6000 AdamW updates (mini-batch 1024, learning rate $0.002$, weight decay $10^{-5}$).
    The backward differentiation formula (BDF) uses relative/absolute tolerances $(10^{-11},10^{-15})$; refinement and independent Radau checks verify the scaling.}}
    \label{fig:learned-cd-time}
\end{figure*}

\begin{revised}
{\it Learned counterdiabatic transport.---}
For transport of the normalized target amplitude $\phi_\lambda=\sqrt{\rho_\lambda}$ itself, the one-sided multiplicative connection is
\begin{equation}
    C_\lambda(x)=\partial_\lambda\log\phi_\lambda(x).
    \label{eq:target-cd-connection}
\end{equation}
It satisfies $C_\lambda\phi_\lambda=\partial_\lambda\phi_\lambda$.
We prescribe $\widehat H_\lambda$ from Eq.~(\ref{eq:score-hamiltonian}) and approximate only this CD connection by $C_\lambda^\theta$.
Let $u\in[0,T]$ be the generation evolution time, distinct from the forward noising time $t$, and let $T>0$ be its total duration.
With $s=u/T\in[0,1]$ and $\lambda'(s)=d\lambda/ds$, the evolution is
\begin{equation}
    \partial_s\chi_s=
    \left[-T\widehat H_{\lambda(s)}
    +\lambda'(s)C^\theta_{\lambda(s)}\right]\chi_s.
    \label{eq:learned-cd-evolution}
\end{equation}
The initial state is $\chi_0=\phi_0$, the normalized state is $\varphi_s=\chi_s/\|\chi_s\|$, and the generated density is $q_T=|\varphi_1|^2$.
When $C^\theta=C$, Eq.~(\ref{eq:learned-cd-evolution}) gives $\chi_s=\phi_{\lambda(s)}$ exactly for every $T$.

To construct $C^\theta$ from a trained score, consider Ornstein--Uhlenbeck noising, $dX_t=-(\beta/2)X_tdt+\sqrt{\beta}\,dW_t$, with rate $\beta>0$, $D=\beta/2$, and standard $d$-dimensional Wiener process $W_t$.
Write $p_t$ for its marginal density on $0\leq t\leq1$, starting from the data density $p_0$, and set $\rho_\lambda=p_{1-\lambda}$.
Here $\lambda=1-t$ reverses the noising coordinate: $\rho_0=p_1$ is the initial noised distribution, whereas $\rho_1=p_0$ is the final data target.
Writing the reverse Fokker--Planck equation as $\partial_\lambda\rho_\lambda=-\nabla\cdot(v_\lambda\rho_\lambda)$ gives the deterministic probability-flow velocity $v_\lambda=b_{\rm gen}-DS_\lambda=(\beta/2)(x+S_\lambda)$.
The continuity equation gives $C_\lambda=-(1/2)\nabla\cdot v_\lambda-(1/2)v_\lambda\cdot S_\lambda$.
Replacing $S_\lambda$ by the fixed neural estimate yields
\begin{equation}
    C^\theta_\lambda
    =-\frac{\beta}{4}\left[
    d+\nabla\cdot S^\theta_\lambda
    +(x+S^\theta_\lambda)\cdot S^\theta_\lambda
    \right],
    \label{eq:learned-cd-score}
\end{equation}
where $S^\theta_\lambda$ is evaluated at forward noising time $1-\lambda$.
For the score error $\epsilon_\lambda=S^\theta_\lambda-S_\lambda$, the CD residual $\delta C_\lambda=C^\theta_\lambda-C_\lambda$ is
\begin{equation}
    \delta C_\lambda
    =-\frac{\beta}{4}\left[
    \nabla\cdot\epsilon_\lambda
    +(x+2S_\lambda)\cdot\epsilon_\lambda
    +|\epsilon_\lambda|^2
    \right].
    \label{eq:cd-learning-residual}
\end{equation}

{\it Boundary cancellation.---}
Writing $e_s=\chi_s-\phi_{\lambda(s)}$ gives the exact deviation equation
\begin{equation}
    \partial_s e_s=
    \left[-T\widehat H_{\lambda(s)}+\lambda'(s)C^\theta_{\lambda(s)}\right]e_s
    +\lambda'(s)\delta C_{\lambda(s)}\phi_{\lambda(s)}.
    \label{eq:cd-error-evolution}
\end{equation}
Here $e_0=0$; the residual drives transverse error, while its parallel component changes normalization.
We use the endpoint-flattened schedule
\begin{equation}
    \lambda_r(s)=1-(1-s)^{r+1}.
    \label{eq:simple-schedule}
\end{equation}
\begin{proposition}[Boundary cancellation of learned-CD transport]
\label{prop:learned-cd-boundary}
Consider Eq.~(\ref{eq:learned-cd-evolution}) on a fixed finite spatial grid with $\chi_0=\phi_0$.
Assume that $\widehat H_\lambda$ is smooth and real self-adjoint, has the unique normalized positive zero mode $\phi_\lambda$, and has all excited eigenvalues bounded below by $\Delta_{\min}>0$ uniformly on $0\leq\lambda\leq1$.
Let $C^\theta_\lambda$ be smooth and independent of $T$.
For fixed integer $r\geq0$, a smooth increasing schedule satisfying $\lambda(0)=0$, $\lambda(1)=1$, and $\lambda^{(j)}(1)=0$ for $j=1,\ldots,r$ gives $H^2(q_T,\rho_1)=O(T^{-2r-2})$ as $T\to\infty$.
For the schedule in Eq.~(\ref{eq:simple-schedule}), more precisely,
\begin{equation}
    H^2(q_T,\rho_1)
    =\frac{A_r}{T^{2r+2}}+o(T^{-2r-2}),
    \label{eq:main-bound}
\end{equation}
where $H^2(q,p)=(1/2)\int(\sqrt q-\sqrt p)^2dx$.
Define $Q_1=1-\ket{\phi_1}\bra{\phi_1}$ and let $\mathcal R_1$ be the inverse of $\widehat H_1$ on its excited subspace, extended by zero on the ground state.
Then
\begin{equation}
    A_r=\frac{[(r+1)!]^2}{2}
    \left\|\mathcal R_1^{r+1}Q_1\delta C_1\phi_1\right\|^2,
    \label{eq:cd-endpoint-prefactor}
\end{equation}
with $\delta C_1=C^\theta_1-C_1$ the terminal CD residual.
\end{proposition}

Equations~(\ref{eq:main-bound}) and (\ref{eq:cd-endpoint-prefactor}) follow by applying the imaginary-time expansion and endpoint analysis of Morita and Nishimori (Secs.~4.2 and 4.4.2)\cite{MoritaNishimori2008MathFoundation} to Eq.~(\ref{eq:cd-error-evolution}) with forcing $\lambda'\delta C\phi$.
Projection onto excited states and repeated integration by parts suppress the initial endpoint exponentially; terminal cancellation leaves an amplitude of order $T^{-r-1}$, yielding the Hellinger coefficient in Eq.~(\ref{eq:cd-endpoint-prefactor}).
Exact CD, $\delta C=0$, gives $e_s=0$ at every $T$.
The schedule reduces the terminal effect of a fixed score error without changing the network's score mean squared error.

{\it Numerical verification.---}
We trained one time-conditioned neural score on the analytic scores of a noised two-dimensional Gaussian mixture.
Training uses uniformly sampled forward noising times and minimizes squared score error with AdamW.
The saved weights are frozen for every $T$ and $r$.
We construct the Hamiltonian analytically from the same density path, so the experiment isolates learning error in the CD term.
The initial amplitude is the square root of the analytic noised density, and the target is the data-density amplitude.

We directly integrated the norm-preserving form of Eq.~(\ref{eq:learned-cd-evolution}) on the full spatial grid.
A ground-state-preserving nearest-neighbor discretization enforces $\widehat H_\lambda\phi_\lambda=0$; grid amplitudes absorb the uniform quadrature weights.
Time-only scalar shifts of the CD connection drop out on normalization.
Deviation coordinates avoid subtractive cancellation at small final errors; direct integration of the unnormalized state with the implicit Runge--Kutta method Radau provides an independent check.
The learned score divergence is evaluated by automatic differentiation, with cubic interpolation in the noising-time coordinate.
We checked tighter tolerances, doubled interpolation resolution, and spatial refinement.
The exact-score CD control follows the prescribed density to numerical precision.

Figure~\ref{fig:learned-cd-time} compares direct evolution with Proposition~\ref{prop:learned-cd-boundary}, varying $T$ independently of integration resolution.
Free-exponent fits reproduce the predicted slopes for the fixed learned connection.
The prefactors agree with Eq.~(\ref{eq:cd-endpoint-prefactor}), computed independently by endpoint linear solves.
Spatial refinement preserves the exponent sequence; endpoint theory and dynamics are compared for each discretization.

{\it Discussion.---}
Terminal cancellation suppresses the distributional effect of a fixed learned-CD residual without retraining.
The asymptotic statement holds for each fixed $r$; its prefactor and crossover time depend on $r$, so increasing the order need not improve every finite-time result monotonically.
In data-driven applications, the forward drift $b_t$ and noising transition density $K_t(x\mid x_0)$ are specified, whereas the population data density $p_\star(x_0)$ is unknown.
Given $N$ independent training samples $x_1,\ldots,x_N$ from $p_\star$, the empirical noised density at $t>0$ is $\widehat p_t(x)=(1/N)\sum_{i=1}^N K_t(x\mid x_i)$.
It estimates the population noised density $p_t(x)=\int K_t(x\mid x_0)p_\star(x_0)\,dx_0$, with $p_0=p_\star$.
In the generation coordinate $\lambda=1-t$, the ideal path is therefore $\rho_\lambda=p_{1-\lambda}$ and its terminal target is $\rho_1=p_0=p_\star$.
The estimated counterpart of $\rho_\lambda$ is obtained by reversing the empirical noising path $\widehat p_t$.
Because the empirical distribution at $t=0$ consists of point masses, we use a fixed normalized, smooth positive regularization of this path, denoted by $\bar\rho_\lambda$.
Its endpoint $\bar\rho_1$ estimates the population target $\rho_1$; it need not equal $\rho_1$ even with exact transport.
Approximating the score of this estimated path by a network introduces an additional CD residual.
Denote by $q_T^{\rm est}(x)$ the normalized density generated after total time $T$ by Eq.~(\ref{eq:learned-cd-evolution}), with the Hamiltonian and approximate CD connection constructed for this estimated path.
Writing $H(q,p)=\sqrt{H^2(q,p)}$ for the Hellinger distance defined below Eq.~(\ref{eq:main-bound}), the triangle inequality gives
\begin{equation}
 H(q_T^{\rm est},\rho_1)\leq H(q_T^{\rm est},\bar\rho_1)+H(\bar\rho_1,\rho_1).
 \label{eq:statistical-transport-split}
\end{equation}
The first term measures transport error relative to the estimated target; the second measures statistical target-estimation error relative to the population.
Under the stated Hamiltonian assumptions, boundary cancellation suppresses the first term without changing the second.
It does not correct the finite-data bias or variance of the estimated target.
The present benchmark removes this statistical uncertainty by prescribing the population density and base Hamiltonian analytically.
Statistical estimation can instead be improved by matching model complexity and noise cutoff to sample size and density regularity, as in minimax analyses of diffusion estimators.\cite{OkoAkiyamaSuzuki2023}
Early stopping of score-network training can also limit memorization while preserving generalization in studied image and language models.\cite{FaveroSclocchiWyart2025}
These estimation procedures act before transport and can complement terminal cancellation, provided the resulting fixed density path satisfies the smoothness and gap assumptions.
\end{revised}

\begin{acknowledgments}
This work was supported by the Cross-ministerial Strategic Innovation Promotion Program (SIP) of the Cabinet Office, Government of Japan (No. 23836436).
\end{acknowledgments}

\bibliographystyle{jpsj}
\bibliography{score_hamiltonian_mn_prl_refs}

\end{document}